\documentclass[10pt,conference]{IEEEtran}
\usepackage{cite}
\usepackage[tight,footnotesize]{subfigure}
\usepackage{graphicx}
\usepackage[cmex10]{amsmath}
\usepackage{amssymb}
\usepackage{mathtools}
\usepackage{multirow}
\usepackage{notoccite}
\usepackage{color} 
\usepackage{epsfig}
\usepackage{latexsym}
\usepackage{subfigure}
\usepackage{pstricks}
\usepackage{pdfpages}
\usepackage{epstopdf}
\DeclareGraphicsExtensions{.eps}
\usepackage{algorithmic}
\usepackage{algorithmic}
\usepackage{algorithm}
\usepackage{multirow}
\usepackage{graphicx}
\usepackage{xurl}

\begin{document}
	\title{Combining Relaying and Reflective Surfaces: Power Consumption and Energy Efficiency Analysis}
	\author{\IEEEauthorblockN{Zaid Abdullah$^\dagger$,  George C. Alexandropoulos$^\ddagger$, Steven Kisseleff$^\dagger$, Symeon Chatzinotas$^\dagger$, and Bj$\ddot{\text{o}}$rn Ottersten$^\dagger$ \\
			\\ $^\dagger$ Interdisciplinary Centre for Security, Reliability and Trust, University of Luxembourg, Luxembourg \\
			$^\ddagger$ Department of Informatics and
Telecommunications, National and Kapodistrian University of Athens, Greece \\
	E-mails: \{zaid.abdullah, steven.kisseleff, symeon.chatzinotas,  bjorn.ottersten\}@uni.lu; alexandg@di.uoa.gr}}
	
	\maketitle
	\begin{abstract}
	Hybrid relaying networks (HRNs) combining both a relay and an intelligent reflective surface (IRS) can lead to enhanced rate performance compared to non-hybrid relaying schemes, where only either an IRS or a relay is utilized.  However, utilizing both the relay and the IRS simultaneously results in higher power consumption for the HRNs compared to their counterpart.  In this work, we study the required transmit power levels and the energy efficiency (EE) of HRNs utilizing both a half-duplex decode-and-forward (HD-DF) relay and an IRS, and compare their performance with non-hybrid relaying schemes. The impact of the required channel estimation overheads is considered when the reflective beamforming design (RBD) at the IRS is carried out under both instantaneous and statistical channel state information models. Also, the investigation is performed for both slow- and fast-changing environments. In terms of the transmit power requirements, our results show that HRNs can lead to higher power savings if the number of reflective elements at the IRS is not very large. However, non-hybrid relaying schemes are shown to be more energy-efficient, unless the targeted rate is high and the IRS is distant from both transmitter and receiver but within a close proximity to the relay.  
	\end{abstract}
	\section{Introduction}
	Sustainability is a core component in modern wireless communications, and for the upcoming sixth generation (6G) systems, an energy efficiency (EE) of up to 1 Terabit/Joule is anticipated \cite{white}. Such ultra-high EE is quite challenging with very large active antenna arrays using a large number (i.e. in the order of hundreds) of power-demanding radio-frequency (RF) chains for their operation.
	\par On the other hand, intelligent reflective surfaces (IRSs) are envisioned to be an attractive solution for energy-efficient communications. IRSs are nearly-passive planar surfaces capable of tweaking the wireless environment by means of smart reflections of impinging signals \cite{jian2022reconfigurable}. Each surface consists of a large number of small unit cells (UCs), which can be digitally configured to introduce phase and/or amplitude manipulations on impinging electromagnetic waves. The UCs at the IRS are nearly passive components, and they do not require power-hungry RF chains to provide signal reflections. 
	\par In principle, the IRS is similar to a multi-antenna amplify-and-forward relay with two main differences. The first one is that IRSs can provide almost instant signal reflections without introducing large delays as it is the case with active relaying, and the second main difference is that IRSs are nearly passive devices that cannot provide active power amplifications, and are therefore highly energy-efficient. For a detailed comparison between relays and IRSs, we refer the reader to the works in \cite{huang2019reconfigurable, bjornson2019intelligent} and the references therein. 
	\par Recently, few works have demonstrated that hybrid relaying networks (HRNs) amalgamating both relays and IRSs can bring about large improvements in terms of achievable rates and/or total transmit powers \cite{abdullah2020hybrid, abdullah2020optimization, Obeed, yildirim2021hybrid}. In particular, the idea of HRNs was first reported in \cite{abdullah2020hybrid}, where it was demonstrated that a cooperative network comprising both an IRS and a single-antenna half-duplex (HD) decode-and-forward (DF) relay can achieve a large rate improvement compared to utilizing only an IRS (i.e. without a relay), given that the number of UCs and/or the transmit power are/is limited.  Furthermore, the work in \cite{kang2021irs} investigated the location and deployment strategy of IRSs in HRNs. Finally, the work in \cite{doubleRIS} investigated the number of relays and transmission strategy for maximum rate performance in double-IRS assisted networks, where the signal is subject to reflections from two spatially separated IRSs. 
	\par In this work, we investigate the performance of hybrid and non-hybrid relaying schemes in terms of required transmit powers and EE performance. Unlike previous works on HRNs \cite{abdullah2020hybrid, abdullah2020optimization, Obeed, yildirim2021hybrid, kang2021irs, doubleRIS}, the required overhead to estimate the channel state information (CSI) is taken into account when the reflective beamforming design (RBD) at the IRS is carried out based on instantaneous CSI (iCSI) as well as statistical CSI (sCSI) models, and for both low- and high-mobility scenarios. For the EE, the power consumption model corresponding to both iCSI- and sCSI-based RBD is formulated, and the EE is evaluated for a wide range of targeted rate thresholds. \par The rest of this paper is organized as follows. In Section~\ref{system model}, we present the system model of different relaying schemes. The achievable rates and IRS optimization are tackled in Section~\ref{RBD}. Transmit powers and EE performance are investigated in Section~\ref{EE}. Numerical evaluations and discussions appear in Section \ref{results}. Concluding remarks are given in Section \ref{conclusions}.
	
	\par \textit{Notations}: Matrices and vectors are represented by boldface uppercase and lowercase letters, respectively. The conjugate, transpose, and Hermitian transpose of a vector $\boldsymbol v$ are denoted by $\boldsymbol v^\ast$, $\boldsymbol v^T$, and $\boldsymbol v^H$, respectively. The ($i,j$)th entry of $\boldsymbol V$ is denoted by $[\boldsymbol V]_{i,j}$, while the $n$th entry of $\boldsymbol v$ is $[\boldsymbol v]_n$. The $N\times N$ identity matrix is $\boldsymbol I_N$, while $\boldsymbol 0_{N}$ and $\boldsymbol 1_N$ are vectors of length $N$ with entries of all $0$'s and $1$'s, respectively. The absolute, expected, and trace operators are expressed as $|\cdot|$, $\mathbb E \{\cdot\}$, and $\text{tr}(\cdot)$, respectively. Moreover, $\angle{(v)}$ denotes the phase of a complex number $v$. Finally, $\boldsymbol V = \text{diag}\{\boldsymbol v\}$ is a diagonal matrix whose diagonal are the elements of $\boldsymbol v$.  
	\section{System Model} \label{system model}
\begin{figure}
        \centering
       \includegraphics[scale=1.75]{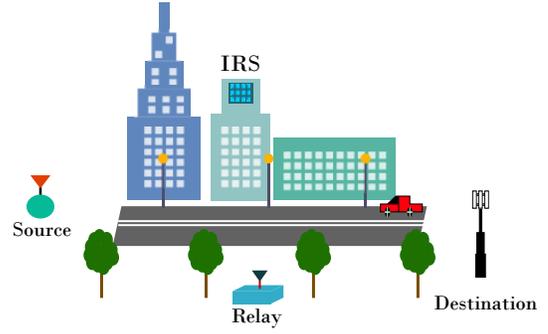} 
        \caption{The considered system in a scattering environment.}
        \label{Fig1}
    \end{figure}

	We consider a network where a source node ($\mathrm S$), aims to transmit data to a destination node ($\mathrm D$), with the help of either an HD-DF relay ($\mathrm R$), an IRS ($\mathrm I$), or both (see Fig. \ref{Fig1}). The source, destination, and relay are each equipped with a single isotropic radiating element, while the IRS has $M$ reflective UCs. A two-dimensional (2D) square array is adopted for the IRS such that $M = M_d^2$ with $M_d$ being the number of UCs per dimension. Spatial correlation at the IRS is taken into account through the correlation matrix $\boldsymbol R$, whose ($n,k$)th entry is \cite{bjornson2020rayleigh}:
	\begin{equation} \label{R} 
		\small 
		\left[\boldsymbol R\right]_{n,k} = \mathrm{sinc} \left( \frac{2\left\|\boldsymbol u_{n} - \boldsymbol u_{k} \right\|}{\lambda}\right), \ \ \ \ \ \ \forall \{n,k\} \in \mathcal M,
	\end{equation}where $\left\|\boldsymbol u_{n} - \boldsymbol u_{k} \right\|$ is the distance between the $n$th and $k$th UCs at the IRS, $\lambda$ is the carrier wavelength, and $\mathcal M = \{1, 2, \ldots, M\}$ is a set containing the indices of all UCs.
	\par Moreover, direct links exist between all nodes except between the source and destination due to blockage, which justifies the deployment of the relay and/or IRS. Block fading is adopted such that the response of channels remains constant within the duration of transmitting one data frame, but changes independently from one frame to another. \par In the following, we formulate the expressions of received
signals and the corresponding signal-to-noise ratios (SNRs)
for the relay-assisted, IRS-assisted, and HRN cases.
	\subsection{Relay-Assisted Scenario}
	In this case, only the relay is utilized to facilitate the communication, which takes place over two phases.
	\subsubsection{First-hop} During this phase, $\mathrm S$ transmits a block of data to $\mathrm R$,\footnote{The indices of different data frames and information symbols are dropped for the sake of simplicity, and without having any impact on the analysis or the results presented in this work.} and the received signal at the latter is given as
	\begin{equation}
	\small 
	    y_{1\mathrm{R}} = \sqrt{P_1} h_{\mathrm{SR}}~ s + w_{1},
	\end{equation}where the subscripts in $y_{1\mathrm R}$ indicate that this is the first phase of the transmission for the relay-assisted scenario, $P_1$ is the transmit power in Watts during the first phase, $s$ is the information symbol satisfying $\mathbb E\{\left|s\right|^2\} = 1$, and $w_1 \sim \mathcal N_{\mathbb C} (0, \sigma^2) $ is the additive white Gaussian noise (AWGN) at the relay. In addition, $h_{\mathrm{SR}}\in\mathbb C$ is the channel coefficient between the source and the relay expressed as $h_{\mathrm{SR}} = \sqrt{\rho_{\mathrm{SR}}} {g}_{\mathrm{SR}}$ with $\rho_{\mathrm{SR}}$ being the channel variance, while $g_{\mathrm{SR}} \sim \mathcal N_{\mathbb C} (0,1)$ accounts for the Rayleigh distributed small-scale flat-fading component. \\ The instantaneous received SNR at the relay is thus given as:
	\begin{equation}
	    \small
	    \gamma_{1\mathrm R} = \frac{P_1}{\sigma^2} \left| h_{\mathrm {SR}}\right|^2.
	\end{equation}

	\subsubsection{Second-hop} Here, $\mathrm R$ re-transmits the signal to $\mathrm D$ after performing decoding and re-encoding on $s$. Assuming a successful decoding at $\mathrm R$, the received signal at $\mathrm D$ during the second transmission phase for the relay-assisted scenario, denoted as $y_{2\mathrm R}$, is given as:
	\begin{equation}
	\small 
	    y_{2\mathrm R} = \sqrt{P_2} h_{\mathrm{RD}}~s + w_2.
	\end{equation}where $P_2$ is the transmit power from $\mathrm R$, $h_{\mathrm{ RD}} = \sqrt{\rho_{\mathrm{RD}}} g_{\mathrm {RD}}$ represents the channel response with $\rho_{\mathrm{RD}}$ being the variance, while $g_{\mathrm {RD}}\sim\mathcal N_{\mathbb C} (0,1)$ accounts for the Rayleigh distributed flat-fading component, and $w_2 \sim \mathcal N_{\mathbb C} (0,\sigma^2)$ is the AWGN at~$\mathrm D$. Therefore, the instantaneous received SNR at $\mathrm D$ is:
	\begin{equation}
	    \small
	    \gamma_{2\mathrm R} = \frac{P_2}{\sigma^2} \left| h_{\mathrm {RD}}\right|^2.
	\end{equation}We next shift our attention to the IRS-assisted case.
	\subsection{IRS-Assisted Scenario}
	In this scenario, only the IRS assists the communication between $\mathrm S$ and $\mathrm D$. A single-phase transmission is sufficient, and the received signal at $\mathrm D$ is:
	\begin{equation}
	    y_{\mathrm {IRS}} = \sqrt{P} (\boldsymbol h_{\mathrm{ID}}^T\boldsymbol \Theta \boldsymbol h_{\mathrm {SI}})~s + w,
	\end{equation}
	where $P$ is the transmit power from the source for the IRS-assisted case, $\boldsymbol h_{\mathrm{ID}} \in \mathbb C^{M\times 1}$ and $\boldsymbol h_{\mathrm{SI}}\in  \mathbb C^{M\times 1}$ are the channel coefficients between $\mathrm I\rightarrow \mathrm D$ and $\mathrm S\rightarrow \mathrm I$. For $x\in\{\mathrm {ID,~ SI}\}$, we have $\boldsymbol h_x = \sqrt{\rho_x} \boldsymbol R ^{\frac{1}{2}}{\boldsymbol g}_x$ with $\rho_x$ being the channel variance between $\{\mathrm S, \mathrm D\}$ and each UC at the IRS, $\boldsymbol R$ is the correlation matrix given in (\ref{R}), and ${\boldsymbol g}_x \sim \mathcal N_{\mathbb C} (\boldsymbol 0_M, \boldsymbol I_M)$ is a vector with independent and identically distributed (i.i.d) Rayleigh fading components. Also, $w\sim\mathcal N_{\mathbb C}(0,\sigma^2)$ is the AWGN at $\mathrm D$, while $\boldsymbol \Theta \in \mathbb C^{M\times M}$ controls the response of each UC at the IRS for the IRS-assisted transmission, and it can be expressed as:
	\begin{equation}
	\small 
    \boldsymbol \Theta = \text{diag}\left\{\left[\mu_{1} e^{\jmath \left[{\boldsymbol {\theta}}\right]_1}, \mu_{2} e^{\jmath \left[ {\boldsymbol {\theta}}\right]_2}, \cdots, \mu_{M} e^{\jmath \left[{\boldsymbol {\theta}}\right]_M}\right]\right\},  
	\end{equation}where $\mu_{m} \in [0,1]$ and $[\boldsymbol \theta]_m \in [0, 2\pi]$ are the reflection amplitude and phase of $m$th ($m\in\mathcal M$) UC at the IRS.\\ The instantaneous received SNR at the destination for IRS-assisted transmission is given as:
	\begin{equation} \label {gamma_IRS}
	    \small 
	    \gamma_{\mathrm{IRS}} = \frac{P}{\sigma^2} \left|\boldsymbol h_{\mathrm{ID}}^T\boldsymbol \Theta \boldsymbol h_{\mathrm {SI}}\right|^2.
	\end{equation}Next, we introduce the system model for the HRN.
	\subsection{HRN scenario}
	In this case, both the HD relay and IRS contribute to the communication between the source and destination \cite{abdullah2020hybrid}, and the data transmission requires two phases.
	\subsubsection{First-hop}
	During this phase, the source transmits its signal to the relay through the direct link and the link via the IRS. The received signal at the relay is given as:  
	{\small \begin{align}
	 y_{1\mathrm H} = \sqrt{P_1} \left(h_{\mathrm{SR}} + \boldsymbol h_{\mathrm{I}\mathrm{R}}^T \boldsymbol \Theta_1 \boldsymbol h_{\mathrm S\mathrm I}\right) s + w_{1},
	\end{align}}where the subscripts in $y_{1\mathrm H}$ reflects the first phase of the HRN scenario, $\boldsymbol h_{\mathrm{IR}} \in \mathbb C^{M\times 1}$ is a vector containing the channel coefficients between $\mathrm I\rightarrow \mathrm R$ given as $\boldsymbol h_{\mathrm{IR}} = \sqrt{\rho_{\mathrm{IR}}} \boldsymbol R ^{\frac{1}{2}}{\boldsymbol g}_{\mathrm{IR}}$ with $\rho_{\mathrm{IR}}$ being the channel variance, and ${\boldsymbol g}_{\mathrm{IR}} \sim \mathcal N_{\mathbb C} (\boldsymbol 0_M, \boldsymbol I_M)$ is the Rayleigh distributed flat-fading channel vector with i.i.d entries. The diagonal matrix $\boldsymbol \Theta_1 \in \mathbb C^{M\times M}$ controls the response of each UC at the IRS during the first phase, such that $\left[\boldsymbol \Theta_1\right]_{m,m} = \mu_{1,m} e^{\jmath [\boldsymbol \theta_1]_m}$, where $\mu_{1,m} \in [0,1]$ and $[\boldsymbol \theta_1]_m \in [0, 2\pi]$ are the reflection amplitude and phase of $m$th ($m\in\mathcal M$) UC at the IRS.\\ The instantaneous received SNR at the relay is given as:
	\begin{equation}
	    \small
	    \gamma_{1\mathrm H} = \frac{P_1}{\sigma^2} \left|h_{\mathrm{SR}} + \boldsymbol h_{\mathrm{I}\mathrm{R}}^T \boldsymbol \Theta_1 \boldsymbol h_{\mathrm S\mathrm I}\right|^2.
	\end{equation}
	\subsubsection{Second-hop} During this phase, the relay broadcasts the signal to the destination through the direct link and reflections from the IRS. The received signal at the destination is:
	{\small \begin{align}
	 y_{2\mathrm H} = \sqrt{P_2} \left(h_{\mathrm{RD}} + \boldsymbol h_{\mathrm{I}\mathrm{D}}^T \boldsymbol \Theta_2 \boldsymbol h_{\mathrm R\mathrm I}\right) s + w_{2},
	\end{align}}where $\boldsymbol h_{\mathrm{RI}} = \sqrt{\rho_{\mathrm{RI}}} \boldsymbol R ^{\frac{1}{2}}{\boldsymbol g}_{\mathrm{RI}} \in \mathbb C^{M\times 1}$ is a vector containing the channel coefficients between $\mathrm R\rightarrow \mathrm I$  with $\rho_{\mathrm{RI}}$ being the channel variance, and ${\boldsymbol g}_{\mathrm{RI}} \sim \mathcal N_{\mathbb C} (\boldsymbol 0_M, \boldsymbol I_M)$ is the Rayleigh distributed flat-fading channel vector with i.i.d entries. Also, $\boldsymbol \Theta_2 \in \mathbb C^{M\times M}$ is the IRS reflection matrix during the second transmission phase, such that $\left[\boldsymbol \Theta_2\right]_{m,m} = \mu_{2,m} e^{\jmath [\boldsymbol \theta_2]_m}$, where $\mu_{2,m} \in [0,1]$ and $[\boldsymbol \theta_2]_m \in [0, 2\pi]$ are the reflection amplitude and phase of $m$th ($m\in\mathcal M$) UC at the IRS during the second transmission phase.\\ The instantaneous received SNR at the destination is:
	\begin{equation}
	    \small
	    \gamma_{2\mathrm H} = \frac{P_2}{\sigma^2} \left|h_{\mathrm{RD}} + \boldsymbol h_{\mathrm{I}\mathrm{D}}^T \boldsymbol \Theta_2 \boldsymbol h_{\mathrm R\mathrm I}\right|^2.
	\end{equation}In the next section, we formulate the achievable rate expressions under both iCSI- and sCSI-based RBD of the IRS. 
	\section{Achievable Rates and RBD} \label{RBD}
	In this section, we formulate the achievable rate expressions for the three different relaying schemes. We take into account the amount of training required to estimate the channels under both iCSI- and sCSI-based RBD.
	\par We denote the length of the coherence interval (in samples) by $\tau_c$, while $L\ge 1$ denotes the number of samples (i.e., pilot signals) utilized to estimate the channel response of a single link.\footnote{Here, we focus on the impact of required training for channel estimation on the rate and hence, the EE performance, while a perfect estimation accuracy with $L$ pilot signals is assumed throughout this work. Nonetheless, the impact of estimation errors on HRNs was investigated in our previous work in \cite{abdullah2020hybrid}.} A frame-based transmission is assumed, where the frame length is aligned with the coherence interval. Furthermore, each frame contains $\tau_p = LT$ pilot symbols with $T$ being the number of channel links to be estimated, followed by $\tau = \tau_c-\tau_p-\tau_g$ data symbols.\footnote{The parameter $\tau_g$ reflects a time gap between pilot and data symbols, which is required only for networks utilizing the IRS. During this period, the receiving node performs the RBD of the IRS, based on the estimated channels, and feedback the optimized phase-shifts to the IRS controller.} 
	\subsection{Relay-Assisted Scenario}
	In this case, and since $\{\mathrm {S, R, D}\}$ are each equipped with a single antenna, only $L$ samples are required to estimate the channel per hop. During the first transmission phase, $\mathrm S$ transmits $L$ pilots to $\mathrm R$ at the start of each coherence interval, and the latter utilizes the received samples to estimate the channel between its antenna and the source (i.e. estimates $h_{\mathrm{SR}}$) and recover the original signal. Similarly, in the second phase, $\mathrm R$ transmits $L$ pilots to $\mathrm D$ for the latter to estimate $h_{\mathrm{RD}}$ and perform the decoding operation. \\ Therefore, the achievable rate for the relay-assisted network with an HD-DF relay is given as:
	\begin{equation} \label{rate_r}
	    \small 
	    \mathcal R_{\mathrm R} = \eta_{\mathrm R} \min \Big\{ \log_2\left(1 + \gamma_{1\mathrm R}\right), \log_2\left(1 + \gamma_{2\mathrm R}\right) \Big\},
	\end{equation}where $\eta_{\mathrm R} = \frac{\tau_c - L}{2\tau_c}$ and the division over two is the result of the HD transmission.
	\subsection{IRS-Assisted Scenario}
	In this case, the RBD at the IRS can be carried out based on either iCSI or sCSI models. In the following, we tackle each case separately. 
	\subsubsection{iCSI-based RBD}
	When the phase shifts at the IRS are reconfigured at each coherence interval, each of the $M$ sub-links of the cascaded channel needs to be estimated (i.e. $[\boldsymbol h_{\mathrm {ID}}]_1 [\boldsymbol h_{\mathrm {SI}}]_1, \cdots, [\boldsymbol h_{\mathrm {ID}}]_M [\boldsymbol h_{\mathrm {SI}}]_M$). As such, at the start of each coherence interval, $LM$ pilots need to be transmitted from $\mathrm S$ to $\mathrm D$ through the IRS to estimate all $M$ channel links at $\mathrm D$, which will then inform the IRS control unit of the optimized phase-shift values through a dedicated control channel.\footnote{Note that it is possible to reduce the amount of training even under iCSI-based RBD by equipping the IRS with estimation capabilites as in  \cite{taha2021enabling}. In addition, strong spatial correlation or channel sparsity can also result in reduced pilot signaling. However, here we focus on a general scenario that does not rely on any channel conditions and/or IRS capabilities.} \\ In such a case, the optimal phase-shift of $m$th UC is $[\bar{\boldsymbol \theta}^\star]_{m} = - \angle\left([\boldsymbol h_{\mathrm{ID}}]_m [\boldsymbol h_{\mathrm{SI} }]_m\right)$ \cite{abdullah2020hybrid}, where the bar notation is used to indicate that the RBD is carried out based on iCSI. Assuming that all UCs have the same reflection amplitude of $\mu$, the maximum received SNR is:
	\begin{equation}
	    \small 
	    \bar{\gamma}_{\mathrm {IRS}} = \frac{P}{\sigma^2}\Big(\mu\sum_{m\in \mathcal M}\Big|[\boldsymbol h_{\mathrm{ID}}]_m [\boldsymbol h_{\mathrm{SI} }]_m\Big|\Big)^2. 
	\end{equation}
	\subsubsection{sCSI-based RBD}
	To reduce the amount of training, one can optimize the response of all UCs based on the statistical CSI model, which is independent of the instantaneous channels' realizations and varies very slowly in practice \cite{demir2022channel}. In this case, the RBD is carried out to maximize the ergodic SNR $\mathbb{E} \left\{\gamma_{\mathrm{IRS}}\right\}$ given in (\ref{gamma_IRS}). Assuming that each UC has a reflection amplitude of $\mu$, an optimal configuration of the IRS can be obtained by setting $\hat{\boldsymbol \Theta}^{\star} = \mu{\boldsymbol I_M}$ (see Appendix A for details), where the hat notation indicates that an sCSI model is adopted for the RBD. Then, the overall cascaded channel $(\boldsymbol h_{\mathrm {ID}}^T \hat{\boldsymbol \Theta}^\star \boldsymbol h_{\mathrm{SI}} = \mu \boldsymbol h_{\mathrm {ID}}^T \boldsymbol h_{\mathrm{SI}})$ can be treated as a single link, and hence, only $L$ pilot samples are required for channel estimation (CE) at the start of each coherence interval. \\ Accordingly, the received SNR under sCSI-based RBD is:
	\begin{equation}
	    \small
	    \hat{\gamma}_{\mathrm{IRS}} = \frac{P}{\sigma^2} \Big| \mu\boldsymbol h_{\mathrm {ID}}^T \boldsymbol h_{\mathrm{SI}}\Big|^2.
	\end{equation}\par It follows that the achievable rate for the IRS-assisted scenario can be expressed as: 
	\begin{equation}
	    \small 
	   {\mathcal R}_{\mathrm{IRS}} = \tilde{\eta}_{\mathrm{IRS}} \log_2\Big(1+\tilde{\gamma}_{\mathrm {IRS}}\Big),
	\end{equation}where $\tilde{\gamma}_{\mathrm{IRS}}\in\{\bar{\gamma}_{\mathrm{IRS}}, \hat{\gamma}_{\mathrm{IRS}}\}$ and $\tilde{\eta}_{\mathrm{IRS}}\in\{\bar{\eta}_{\mathrm{IRS}}, \hat{\eta}_{\mathrm{IRS}}\}$, depending on whether the RBD is carried out based on iCSI or sCSI. Regarding the parameter $\tilde{\eta}_{\mathrm{IRS}}$, we have $\bar{\eta}_{\mathrm{IRS}} = \frac{\tau_c - LM - \tau_g}{\tau_c}$, while $\hat{\eta}_{\mathrm{IRS}} = \frac{\tau_c - L}{\tau_c}$.
	\subsection{HRN Scenario}
	For the HRN, one can also adopt either the iCSI or the sCSI based RBD. In each case, the CE is performed at $\mathrm R$ and $\mathrm D$, during the first and the second hops, respectively.  
	\subsubsection{iCSI-based RBD}
	In this case, and for each of the two transmission phases, the $M$ links through the IRS as well as the direct link need to be estimated. Therefore,  $LM+L$ pilot samples are required at the start of each coherence interval. 
	\par The phase-shifts are adjusted to maximize the instantaneous SNRs at $\mathrm R$ and $\mathrm D$, during the first and second transmission phases, respectively. The optimal phase response of the $m$th UC ($m\in\mathcal M$) during the first phase is $[\bar{\boldsymbol \theta}^\star_1]_{m} = \angle({h_\mathrm{SR}}) - \angle\left([\boldsymbol h_{\mathrm{IR}}]_m [\boldsymbol h_{\mathrm{SI} }]_m\right)$, while during the second phase of transmission we have $[\bar{\boldsymbol \theta}^\star_2]_{m} = \angle({h_\mathrm{RD}}) - \angle\left([\boldsymbol h_{\mathrm{ID}}]_m [\boldsymbol h_{\mathrm{RI} }]_m\right)$. Assuming a fixed reflection amplitude of $\mu$ at each UC, the corresponding SNRs at $\mathrm R$ and $\mathrm D$ are expressed as follows \cite{abdullah2020hybrid}
\begin{subequations}
\begin{equation}
\small 
    \bar{\gamma}_{1\mathrm H} = \frac{P_1}{\sigma^2} \Big( \left|h_{\mathrm SR}\right| + \mu \sum_{m\in \mathcal M}\left|\left[\boldsymbol h_{\mathrm {IR}}\right]_m \left[\boldsymbol h_{\mathrm {SI}}\right]_m  \right| \Big)^2,
\end{equation}
\begin{equation}
\small 
        \bar{\gamma}_{2\mathrm H} = \frac{P_2}{\sigma^2} \Big( \left|h_{\mathrm RD}\right| + \mu \sum_{m\in \mathcal M}\left|\left[\boldsymbol h_{\mathrm {ID}}\right]_m \left[\boldsymbol h_{\mathrm {RI}}\right]_m  \right| \Big)^2.
\end{equation}
\end{subequations}
\subsubsection{sCSI-based RBD} Here, the phase optimization is carried out to maximize the ergodic SNRs, and the optimal RBD during both transmission phases can be obtained as $\hat{\boldsymbol \Theta}_i = \mu \boldsymbol I_M~(i\in\{1,2\})$ (see Appendix B for details). The overall effective channel between $\mathrm S$ and $\mathrm R$ $\small \left(h_{\mathrm{SR}} + \mu \boldsymbol h_{\mathrm{I}\mathrm{R}}^T \boldsymbol h_{\mathrm S\mathrm I}\right)$ is treated as a single channel link, and the same applies for the second transmission phase between $\mathrm R$ and $\mathrm D$. Therefore, one only needs $L$ pilot samples per transmission phase to estimate the overall channel link at the start of each coherence interval. 
\\ The corresponding SNRs under sCSI-based RBD are:
\begin{subequations}
\begin{equation}
\small 
    \hat{\gamma}_{1\mathrm H} = \frac{P_1}{\sigma^2} \Big|h_{\mathrm SR} + \mu \boldsymbol h_{\mathrm{I}\mathrm{R}}^T \boldsymbol h_{\mathrm S\mathrm I} \Big|^2,
\end{equation}
\begin{equation}
\small 
    \hat{\gamma}_{2\mathrm H} = \frac{P_2}{\sigma^2} \Big|h_{\mathrm RD} + \mu \boldsymbol h_{\mathrm{I}\mathrm{D}}^T \boldsymbol h_{\mathrm R\mathrm I} \Big|^2.
\end{equation}
\end{subequations}
\par Therefore, the achievable rate of the HRN with an HD-DF relay can be given as:
	\begin{equation} \label{rate_H}
	    \small 
	    {\mathcal R}_{\mathrm H} = \tilde{\eta}_{\mathrm H} \min \Big\{ \log_2\left(1 + \tilde{\gamma}_{1\mathrm H}\right), \log_2\left(1 + \tilde{\gamma}_{2\mathrm H}\right) \Big\},
	\end{equation}where $\tilde{\gamma}_{i\mathrm H}\in\{\bar{\gamma}_{i\mathrm H}, \hat{\gamma}_{i\mathrm H}\}$ ($i\in\{1,2\}$) and $\tilde{\eta}_{\mathrm H}\in\{\bar{\eta}_{\mathrm H}, \hat{\eta}_{\mathrm H}\}$, such that $\bar{\eta}_{\mathrm H} = \frac{\tau_c - (LM+L)-\tau_g}{2\tau_c}$ and $\hat{\eta}_{\mathrm H} = \frac{\tau_c - L}{2\tau_c}$.
	\section{Transmit Power Levels and EE Performance} \label{EE}
Here, we study the EE performance of the different relaying schemes. First, we obtain the minimum required transmit powers to achieve a given data rate threshold of $R_{{th}}$ at the destination. Then, the EE is evaluated based on the required power levels. In general, the EE can be defined as follows \cite{bjornson2019intelligent}:
\begin{equation}
\small 
    \mathrm{EE} = \frac{R_{{th}}} {{P}_{\text{total}}/B},
\end{equation}where $B$ is the bandwidth and $P_{\text{total}}$ is the total power consumption of the considered communication network.
\subsection{Relay-Assisted Scenario}
Let us define $\beta_{\mathrm{SR}} = |h_{\mathrm{SR}}|^2$ and $\beta_{\mathrm{RD}} = |h_{\mathrm{RD}}|^2$. Then, under a transmit power constraint of ${\small P = \frac{P_1 + P_2}{2}}$, it follows that the maximum achievable rate, which is obtained when $\gamma_{1\mathrm R} = \gamma_{2\mathrm R}$ (see Eq.(\ref{rate_r})), is ${\small \mathcal R_{\mathrm R}^\star = \eta_{\mathrm R}\log_2\left(1 + \frac{2P\beta_{\mathrm{SR}} \beta_{\mathrm{RD}} }{(\beta_{\mathrm{SR}} + \beta_{\mathrm{RD}})\sigma^2}\right)}$. Therefore, the required transmit power for the relay-assisted scenario to achieve a rate of $R_{{th}}$ is:
\begin{equation}
\small 
    P_{\mathrm R} = \Big(2^{\frac{R_{{th}}}{\eta_{\mathrm R}}}-1\Big) \frac{\left(\beta_{\mathrm{SR}} + \beta_{\mathrm{RD}}\right)\sigma^2 }{2\beta_{\mathrm{SR}} \beta_{\mathrm{RD}}},
\end{equation}and the total power consumption for the relay-assisted case is:
\begin{equation} \label{P_R}
    \small 
    P_{\text{total}}^{\mathrm R} = \frac{P_{\mathrm R}}{\zeta} + \frac{1}{2} \overline{p}_{\mathrm S} + \frac{1}{2} \overline{p}_{\mathrm D} + \overline{p}_{\mathrm R},
\end{equation}where $\zeta \in (0,1]$ is the power amplifier efficiency, while $\overline{p}_{\mathrm S},~ \overline{p}_{\mathrm R}$, and $\overline{p}_{\mathrm D}$ are the hardware-dissipated power at the source, relay, and destination, respectively. The division over two of $\{\overline{p}_{\mathrm S}, \overline{p}_{\mathrm D}\}$ in (\ref{P_R}) is due to the fact that the source and destination are only active for half of the transmission time. 
\subsection{IRS-Assisted Scenario} Let the channel gains under iCSI- and sCSI-based RBD for the IRS case be $\small \bar{\beta}_{\mathrm{IRS}}=\left(\mu\sum_{m\in \mathcal M}\Big|[\boldsymbol h_{\mathrm{ID}}]_m [\boldsymbol h_{\mathrm{SI} }]_m\Big|\right)^2$ and $\small \hat{\beta}_{\mathrm{IRS}} = \left|\mu \boldsymbol h_{\mathrm{ID}}^T \boldsymbol h_{\mathrm{SI}}\right|^2$, respectively. Then, the required transmit power to achieve a rate of $R_{{th}}$ is:
\begin{equation}
\small 
    {P}_{\mathrm{IRS}} = \left(2^{\frac{R_{{th}}}{\tilde{\eta}_{\mathrm{IRS}}}}-1\right) \frac{\sigma^2}{\tilde{\beta}_{\mathrm{IRS}}}
\end{equation}with $\tilde{\beta}_{\mathrm{IRS}} \in \{\bar{\beta}_{\mathrm{IRS}}, \hat{\beta}_{\mathrm{IRS}}\}$, depending on the RBD criterion.
\par The total power consumption of the IRS case is given as:
\begin{equation}
\small 
    P_{\text{total}}^{\mathrm{IRS}} = \frac{P_{\mathrm{IRS}}}{\zeta} + \overline{p}_{\mathrm S} + \overline{p}_{\mathrm D} + M\left(\overline p_{\text{st}} + \overline{p}_{\text{dyn}}\right)
\end{equation}with $\overline p_{\text{st}}$ and $\overline{p}_{\text{dyn}}$ being the static and dynamic power dissipation at the IRS, respectively. In particular, $\overline p_{\text{st}}$ is the static power that the IRS consumes just for being connected to an energy source, while $\overline p_{\text{dyn}}$ is the power consumed due to the reconfiguration of UCs \cite{report}. Note that when sCSI-based RBD is carried out, $\overline p_{\text{dyn}}$ is equal to zero since the UCs are not reconfigured at each coherence interval. 
\subsection{HRN Scenario}
We define ${\small \bar{\beta}_{1\mathrm H} =  \left( \left|h_{\mathrm SR}\right| + \mu \sum_{m\in \mathcal M}\left|\left[\boldsymbol h_{\mathrm {IR}}\right]_m \left[\boldsymbol h_{\mathrm {SI}}\right]_m  \right| \right)^2}$ and ${\small \bar \beta_{2\mathrm H} =  \left( \left|h_{\mathrm RD}\right| + \mu \sum_{m\mathcal M}\left|\left[\boldsymbol h_{\mathrm {ID}}\right]_m \left[\boldsymbol h_{\mathrm {RI}}\right]_m  \right| \right)^2}$ as the effective channel gains during the first and second transmission phases for the HRN with iCSI-based RBD. Similarly, under sCSI-based RBD, we define ${\small \hat \beta_{1\mathrm H} =  \left|h_{\mathrm SR} + \mu \boldsymbol h_{\mathrm {IR}}^T \boldsymbol h_{\mathrm {SI}} \right|^2}$ and ${\small \hat \beta_{2\mathrm H} = \left|h_{\mathrm RD} + \mu \boldsymbol h_{\mathrm {ID}}^T \boldsymbol h_{\mathrm {RI}} \right|^2}$. Then, to achieve optimal received SNRs, one should optimize the transmit powers such that $\tilde \gamma_{1\mathrm H} = \tilde \gamma_{2\mathrm H}$ (see Eq. (\ref{rate_H})). Therefore, under a transmit power constraint of $P = \frac{P_1 + P_2}{2}$, the maximum achievable rate with optimal transmit powers for the HRN is ${\small \mathcal R_{\mathrm H}^\star = \tilde \eta_{\mathrm H} \log_2\left(1 + \frac{2P\tilde\beta_{1\mathrm H}\tilde\beta_{2\mathrm H}}{(\tilde\beta_{1\mathrm H} + \tilde\beta_{2\mathrm H})\sigma^2}\right)}$, where $\tilde \beta_{i\mathrm H} \in\{\bar \beta_{i\mathrm H}, \hat \beta_{i\mathrm H}\}$ and $i\in\{1, 2\}$.
\\ To achieve a rate of $R_{{th}}$, the required transmit power is:
\begin{equation}
    \small 
    P_{\mathrm H} = \left( 2^{ \frac{R_{{th}}}{\tilde \eta_{\mathrm H}} }  -1\right) \frac{(\tilde \beta_{1\mathrm H} + \tilde \beta_{2\mathrm H})\sigma^2}{2\tilde \beta_{1\mathrm H}\tilde \beta_{2\mathrm H}},
\end{equation}and the total power consumed in such a case is given as:
\begin{equation}
    \small 
    P_{\text{total}}^{\mathrm H} = \frac{P_{\mathrm H}}{\zeta} + \frac{1}{2} \overline{p}_{\mathrm S} + \frac{1}{2} \overline{p}_{\mathrm D} + \overline{p}_{\mathrm R} + M\left(\overline p_{\text{st}} + \overline{p}_{\text{dyn}}\right).
\end{equation}
\begin{figure*}[t]
\centering
\subfigure[\footnotesize $\tau_c = 10^4$ samples, ${R}_{th} = 3$ bits/s/Hz.]{%
\label{Fig2a}%
\includegraphics[width=.311\linewidth]{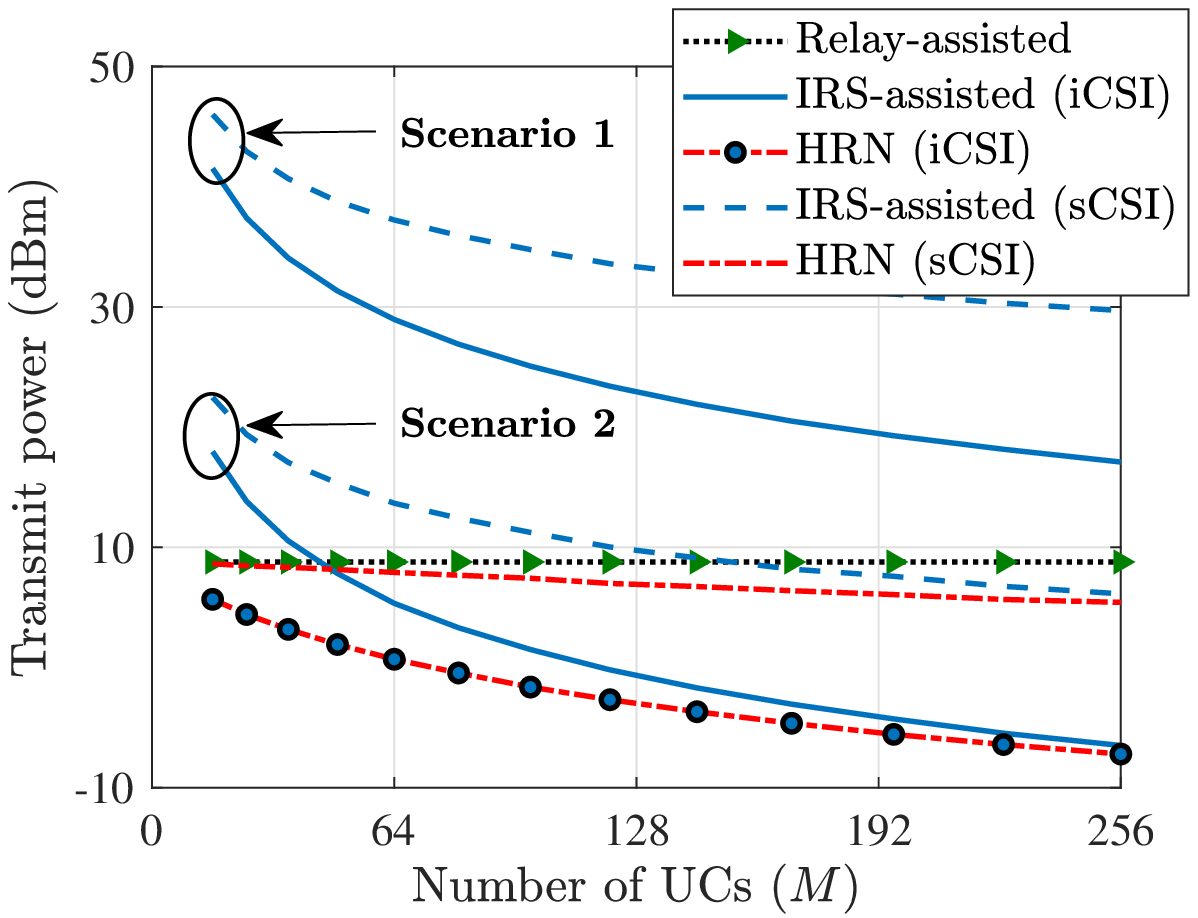}
}\hspace{0.31cm}
\subfigure[\footnotesize $\tau_c = 10^3$ samples, ${R}_{th} = 3$ bits/s/Hz.]{%
\label{Fig2b}%
\includegraphics[width=.311\linewidth]{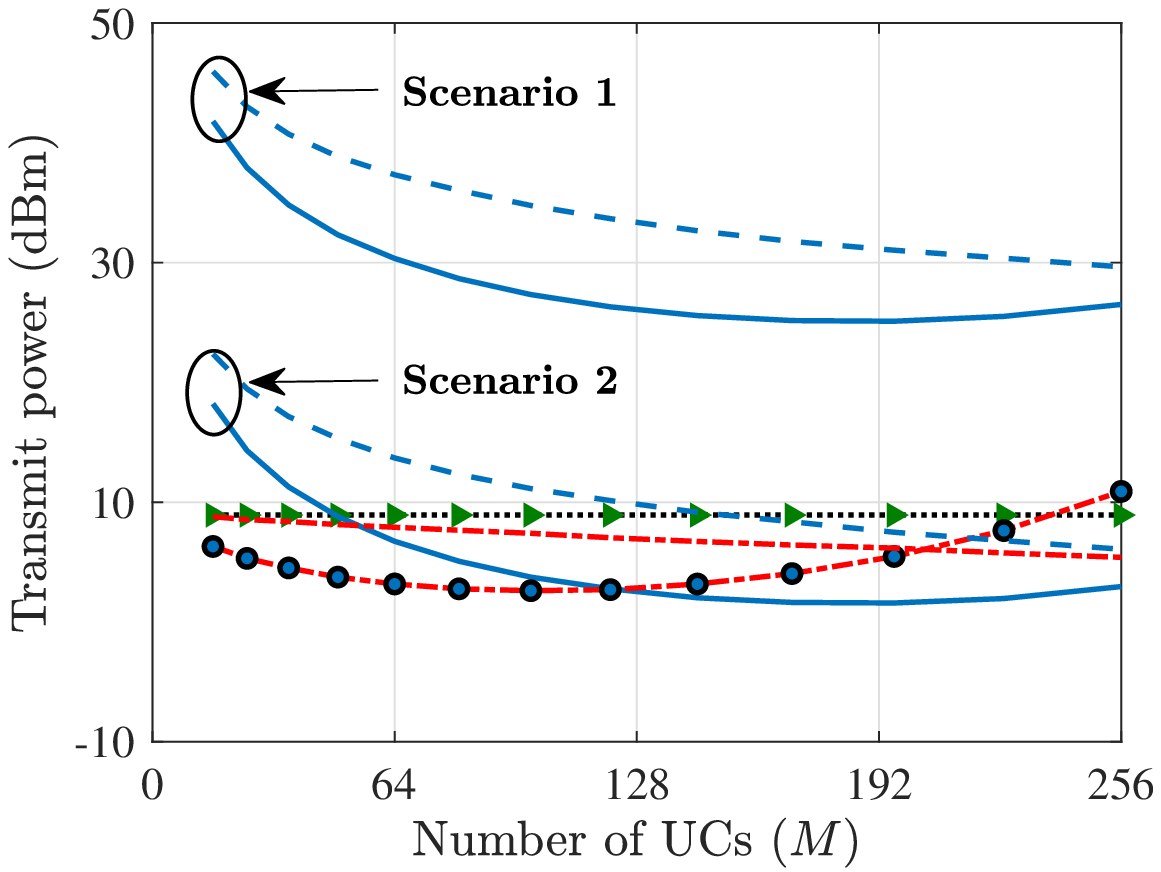}
}\hspace{0.31cm}
\subfigure[\footnotesize $\tau_c = 10^4$ samples, $M = 144$.]{%
\label{Fig2c}%
\includegraphics[width=.311\linewidth]{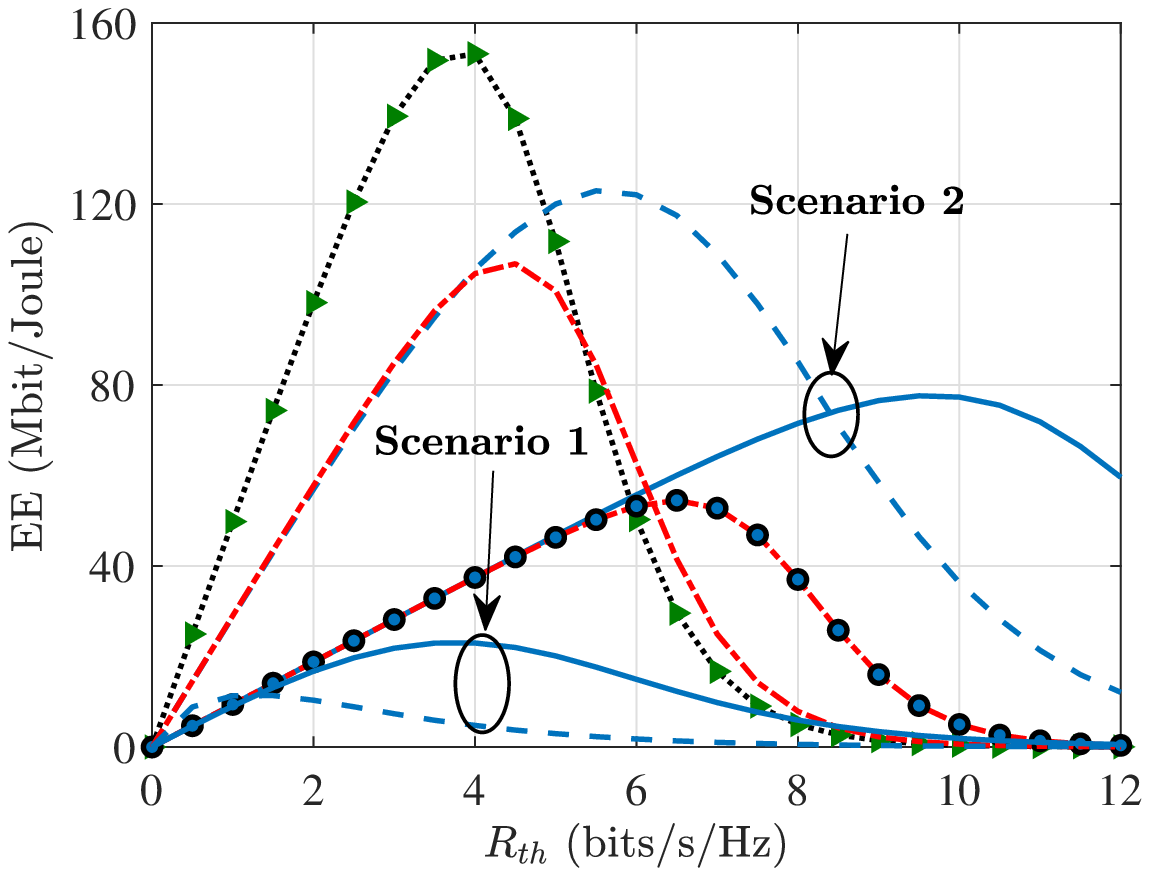}
}%
\caption{Performance comparison among different relaying schemes. The legend in (a) applies to all figures.}\label{fig:exp1}
\end{figure*}
\section{Results and Discussions} \label{results}
We start by introducing the locations of different communication nodes. In particular, a three-dimensional network setup was considered, with the $xyz-$coordinates of the source, relay, and destination being fixed at $(0,~0,~0), ~(100,~0,~0)$, and $(200,~0,~0)$, respectively, all in meters. Regarding the location of the IRS,\footnote{We highlight that for IRS-assisted scenarios, higher channel gains are obtained when the IRS is closest to the source or destination as in such cases the double path-loss is minimal. In contrast, when dealing with HRNs, the IRS should ideally be in a close proximity to the relay\cite{abdullah2020hybrid, abdullah2020optimization, kang2021irs}, and it is well known that under identical channel characteristics of the two sub-links, the relay provides the highest performance enhancement when located in the middle between the two end nodes of the network.} it was located near the relay at $(100,~2,~8)$~meters when dealing with an HRN, while for the IRS-assisted system, we evaluate the performance under two different scenarios. \textbf{Scenario~1}: The IRS is located between the two end nodes at $(100,~2,~8)$~meters similar to the HRN case, and \textbf{Scenario~2}: The IRS is located near the source at $(0,~2,~8)$~meters. 
\par The channel variance between any two nodes $i$ and $j$ was modelled as $\rho_{ij}~[\mathrm{dB}] = 10\log_{10} (\frac{d_{ij}}{d_0})^{-\alpha} - 20$, where $d_{ij}$ is the distance between the two nodes, $d_0 = 1~$m is the reference distance, and $\alpha$ is the path-loss exponent. In particular, a path-loss exponent of $3$ was set for all links that involve the IRS, while a path-loss exponent of $3.7$ was selected for channel links between the relay and both the source and destination.\footnote{The justification of different path-losses is that IRSs can be mounted on the facades of tall buildings, and thereby experiencing a better link quality than relays who can be cooperative users in a dense urban network.}  In addition, the carrier frequency is $1.9$~GHz \cite{demir2022channel}, $\sigma^2 = -107~\mathrm{dBm}$, $L = 1$, $\tau_g = M$, $\mu = 0.9$, $B = 10$~MHz, $\zeta = 0.5$, $\overline{p}_{\mathrm S} = \overline{p}_{\mathrm D} = \overline{p}_{\mathrm R} = 100$~mW \cite{bjornson2019intelligent}, $\overline{p}_{\text{dyn}} = 5$~mW \cite{bjornson2019intelligent}, and $\overline{p}_{\text{st}}=1$~mW. Finally, the IRS element spacing (i.e. the distance between two adjacent UCs located on the same row/column of the IRS surface) is $\lambda/8$.
\par Fig.~\ref{Fig2a} illustrates the required transmit powers of different relaying schemes to achieve a rate threshold of $3$ bits/s/Hz with $\tau_c = 10^4$ samples, which is a typical value for low-mobility scenarios \cite{demir2022channel}. The results show that an HRN with iCSI-based RBD can achieve superior power savings compared to all other schemes. As the number of UCs at the IRS increases, the power savings of the HRN become larger compared to the relay-assisted case, while the opposite holds when it comes to the comparison with the IRS-assisted cases. This is in line with the findings of \cite{abdullah2020hybrid} which stated that an IRS with a very large number of UCs outperforms an HRN with both an HD relay and an IRS. Moreover, for an IRS-assisted network, if the IRS is located within a close proximity of the source, higher power savings can be achieved. This is due to the fact that the double path-loss of an IRS-assisted network worsen as the location of the IRS moves toward the middle point between the two transceiving end nodes.  
\par Fig. \ref{Fig2b} presents a comparison of the transmit powers under a fast-changing environment with $\tau_c = 1000$ samples. The size of the IRS plays a crucial role in such scenarios if the RBD is carried out based on the iCSI. In particular, the achievable rate of an IRS-assisted network could suffer from a rate penalty of $\bar{\eta}_{\text{IRS}}\approx 0.5$ for $M = 256$, which means that about $50\%$ of the transmission time is allocated for the CE and RBD phase. Such a challenge becomes even worse for the HRN with an HD relay, as for the same number of $256$ UCs, the parameter $\bar{\eta}_{\text{H}}\approx 0.25$ means that about $75\%$ of the achievable rate is wasted as a result of the CE with RBD phase and also the HD operation mode. Therefore, and as shown in Fig. \ref{Fig2b}, the HRN with iCSI-based RBD is the most affected when dealing with large IRSs. On the contrary, the sCSI-based RBD cases show improved performance as the number of UCs increases. This is due to the fact that the amount of pilot samples required do not increase with the number of UCs when sCSI is adopted to carry out the IRS-phase configuration. 
\par Finally, Fig. \ref{Fig2c} demonstrates the EE performance. We observe that when the targeted rate is low, the relay-assisted case is by far the most efficient choice. In contrast, for high targeted rates, the HRN is the most energy-efficient system if the IRS was located near the relay, while an optimally placed IRS shows higher efficiency compared to the HRN. Interestingly, an IRS with sCSI-based RBD achieves the highest EE at medium targeted rate thresholds (between $5.3$ and $7.5$ bits/s/Hz), which demonstrates that although the sCSI-based RBD requires higher transmit powers compared to iCSI-based RBD (as shown in Fig.~\ref{Fig2a} and Fig.~\ref{Fig2b}), it can be more energy efficient due to a lower overall power consumption. 
\par From all the above, one can identify the scenarios where HRNs can be efficiently utilized. For example, if the transmit power of a communication device is limited (such as a mobile user or even an IoT device), then HRNs can help achieving higher power savings compared to non-hybrid schemes, while ensuring a targeted rate threshold as shown in Fig. \ref{Fig2a}. Also, when both the transmitter and receiver are far away from the IRS, then incorporating a cooperative relay device that is close to the IRS can lead to better EE when the targeted rate is high as shown in Fig.~\ref{Fig2c}.

\section{Concluding Remarks} \label{conclusions}
We thoroughly investigated the power requirements and EE performance of hybrid and non-hybrid relaying networks under both iCSI- and sCSI-based RBD models. We highlighted the role of various parameters on the power and EE performance, such as the RBD models, effects of high mobility, the number of UCs, the targeted rate, as well as the IRS placement.  
\section*{Acknowledgment}
	This work was supported by the Luxembourg National Research Fund (FNR) through the CORE Project under Grant RISOTTI C20/IS/14773976.
\section*{Appendix A}
The ergodic SNR in (\ref{gamma_IRS}) with correlated Rayleigh fading is:
{\small \begin{eqnarray}
			\mathbb E\{\gamma_{\mathrm{IRS}}\} & = & \frac{P}{\sigma^2} \mathbb E \Big\{\boldsymbol h_{\mathrm {ID}}^T\boldsymbol \Theta \boldsymbol h_{\mathrm {SI}}\boldsymbol h_{\mathrm {SI}}^H\boldsymbol \Theta^H \boldsymbol h_{\mathrm {ID}}^\ast\Big\} \nonumber \\ & = & \frac{P}{\sigma^2} \mathbb E \Big\{\boldsymbol h_{\mathrm {ID}}^T\boldsymbol \Theta \mathbb E\{\boldsymbol h_{\mathrm {SI}}\boldsymbol h_{\mathrm {SI}}^H\} \boldsymbol \Theta^H \boldsymbol h_{\mathrm {ID}}^\ast\Big\} \nonumber \\ & = & \frac{P}{\sigma^2} \rho_{\mathrm{SI}} \mathbb E \Big\{\boldsymbol h_{\mathrm{ID}}^T\boldsymbol \Theta \boldsymbol R \boldsymbol \Theta^H \boldsymbol h_{\mathrm{ID}}^\ast\Big\} \nonumber \\ & = & \frac{P}{\sigma^2} \rho_{\mathrm{SI}} \text{tr}\Big(\mathbb E \Big\{\boldsymbol h_{\mathrm{ID}}^\ast\boldsymbol h_{\mathrm{ID}}^T \Big\} \boldsymbol \Theta \boldsymbol R \boldsymbol \Theta^H \Big) \nonumber \\ & = & \frac{P}{\sigma^2} \rho_{\mathrm{ID}} \rho_{\mathrm{SI}} \text{tr}\big(\boldsymbol R \boldsymbol \Theta \boldsymbol R \boldsymbol \Theta^H \big). 
\end{eqnarray}}Then, from \cite[Theorem 2]{abdullah2022impact}, the optimal solution must satisfy $\boldsymbol \Theta^\star = \text{diag}\{\exp\left(\jmath c\boldsymbol 1_M\right)\}$, with $c$ being any real number. For a reflection amplitude of $\mu$, and by letting $c = 0$, we obtain $\boldsymbol \Theta^\star = \mu\boldsymbol I_M$ as an optimal solution under sCSI-based RBD.  
\section*{Appendix B}
The ergodic SNR for HRN during the first phase is:
{\small \begin{eqnarray}
	    \mathbb E\left\{\gamma_{1\mathrm H}\right\} & = & \mathbb E \left\{\frac{P_1}{\sigma^2}   \left|h_{\mathrm{SR}} + \boldsymbol h_{\mathrm{I}\mathrm{R}}^T \boldsymbol \Theta_1 \boldsymbol h_{\mathrm S\mathrm I}\right|^2 \right\}\nonumber \\ & \stackrel{\mathrm{a}}{=} & \frac{P_1}{\sigma^2} \Big(\mathbb E \left\{\left|h_{\mathrm{SR}} \right|^2 \right\} + \mathbb E \Big\{\left| \boldsymbol h_{\mathrm{I}\mathrm{R}}^T \boldsymbol \Theta_1 \boldsymbol h_{\mathrm S\mathrm I}\right|^2 \Big\} \Big) \nonumber \\ & = & \frac{P_1}{\sigma^2} \left(\rho_{\mathrm{SR}} + \rho_{\mathrm{IR}}\rho_{\mathrm{SI}} \text{tr}\big(\boldsymbol R \boldsymbol \Theta_1 \boldsymbol R \boldsymbol \Theta^H_1 \big)\right),
\end{eqnarray}}where equality ($\mathrm a$) holds due to the statistical independence of the direct and reflected channels. Then, from \cite[Theorem~2]{abdullah2022impact}, and for a reflection amplitude of $\mu$, we obtain the solution $\boldsymbol \Theta_1^\star = \mu\boldsymbol I_M$. The phase-shift matrix during the second transmission phase can be obtained following similar steps.

\bibliographystyle{IEEEtran}
\bibliography{EE_HRNs}	
\end{document}